# Magnetotransport due to conductivity fluctuations in non-magnetic ZrTe$_2$ nanoplates

Jie Wang[1,2*], Yihao Wang[1*], Min Wu[1*‡], Junbo Li[1,2], Shaopeng Miao[1,2], Qingyi Hou[1,2], Yun Li[1,3], Jianhui Zhou[1], Xiangde Zhu[1], Yimin Xiong[1‡], Wei Ning[1‡], and Mingliang Tian[1,4]

[1]*Anhui Key Laboratory of Condensed Matter Physics at Extreme Conditions, High Magnetic Field Laboratory, HFIPS, Anhui, Chinese Academy of Sciences, Hefei 230031, P. R. China.*

[2]*Department of physics, University of Science and Technology of China. Hefei 230026, P. R. China.*

[3]*Department of Materials Science and Engineering, and ARC Centre of Excellence in Future Low-Energy Electronics Technologies (FLEET), Monash University, Clayton, Victoria 3800, Australia.*

[4]*Department of Physics, School of Physics and Materials Science, Anhui University, Hefei 230601, P. R. China.*

*Those authors contribute equally to this work.

‡ To whom correspondence should be addressed. E-mail: minwu@hmfl.ac.cn, yxiong@hmfl.ac.cn, ningwei@hmfl.ac.cn.

Transition metal dichalcogenides with nontrivial band structures exhibit various fascinating physical properties and have sparked intensively research interest. Here, we performed systematic magnetotransport measurements on mechanical exfoliation prepared ZrTe$_2$ nanoplates. We revealed that the negative longitudinal magnetoresistivity observed at high field region in the presence of parallel electric and magnetic fields could stem from the conductivity fluctuations due to the excess Zr in the nanoplates. In addition, the parametric plot, the planar Hall resistivity as function of the in-plane anisotropic magnetoresistivity, has an ellipse-shaped pattern with shifted orbital center, which further strengthen the evidence for the conductivity fluctuations. Our work provides some useful insights into transport phenomena in topological materials.

Transition metal dichalcogenides (TMDCs), typical layered materials with strong bonding within the plane but weak interlayer van der Waals interactions, have attracted great attentions due to versatile electrical, optical, and chemical properties.[1-4] More interestingly, TMDCs are making it possible to tune the physical properties by ion insertion, ionic gating or Moiré method,[5-10] opening up possibilities for both fundamental research and future applications in heterostructure transistor, valleytronics.[11,12] Among the TMDCs, layered zirconium telluride compounds have been drawn intensively research interest because of the non-trivial electron structure and the novel transport properties. For example, many exotic transport phenomena have been observed in three-dimensional (3D) topological material $ZrTe_5$, such as negative longitudinal magnetoresistivity (LMR),[13-15] 3D quantum Hall effect,[16,17] log-periodic oscillations,[18] and band inversion induced four-fold splitting of the non-zero Landau levels.[19] Another zirconium telluride compound that possesses non-trivial band topology is $ZrTe_2$.[20-25] Theoretical studies predicted that $ZrTe_2$ is a topological crystalline insulator protected by crystalline symmetry.[20-22] However, recent angle resolved photoemission spectroscopy (ARPES) studies on both the thin films grown by molecular beam epitaxy and the bulk single crystals reveal that $ZrTe_2$ is a topological semimetal with approximately equal electron and hole carrier densities.[23,25] The magnetotransport measurements on $ZrTe_2$ thin films grown by pulsed-laser deposition (PLD) further presented an indication of topological semimetal feature.[24] Experimentally, the negative LMR had been observed in both thin films and crystals,[24,26] however, the origin is still lacking.

To reveal the underlying physics of the negative LMR, we carried out comprehensive magnetotransport measurements on ZrTe$_2$ nanoplates prepared by mechanical exfoliation. Compared to an unsaturated positive LMR under the applied magnetic field perpendicular to the sample plane, a negative LMR is observed for the in-plane magnetic field parallel and perpendicular to the electric current. By carefully analyzing the feature of negative LMR and the temperature dependent carrier density and mobility, the negative LMR observed at low fields and high fields can be ascribed to the Kondo effect and conductivity fluctuations, respectively. Furthermore, the measured planar Hall effect (PHE) and in-plane anisotropic LMR further supported the existence of conductivity fluctuations due to the excess Zr in ZrTe$_2$ nanoplates.

The ZrTe$_2$ nanoplates were mechanically exfoliated onto 285 nm thick SiO$_2$/Si substrates in Ar-filled glovebox from the bulk crystals grown by the chemical vapor transport method.[26] These nanoplates were then fabricated into standard Hall-bar devices by standard electron-beam lithography and lift-off techniques. To avoid degradation or oxidation during the transport measurements, the devices were covered with poly-methyl methacrylate layers before the transport measurements. Figure 1(a) shows the temperature dependence of resistivity under zero magnetic field for two samples #1 and #2, with thickness of about 100 and 75 nm determined by atomic force microscope, respectively. At the temperatures between 6 and 300 K, the $\rho - T$ curves can be well captured by the Bloch-Grüneisen (BG) model, as indicated by the red curves, suggesting the resistivity at this temperature range ($T > 6$ K) is mainly dominated by the electron-phonon scattering. Remarkably, a resistivity upturn appeared below 6 K, which manifests as logarithmic scale behavior, as shown in the inset of Fig. 1(a). This behavior has also been observed in bulk crystals and was attributed to the Kondo effect.[26] It should be noted that, the residual resistivity ratio [ RRR = $\rho(300K)/\rho(2K)$] in the nanoplates is much smaller than the thin films prepared by PLD,[24] which may due to the existence of excess Zr in the bulk crystals that we used for exfoliation.[26] The atomic ratio in the nanoplates is measured by the energy-dispersive X-ray spectrometry. As shown in Table 1, the excess Zr concentrations in the nanoplates are nearly same with bulk crystals, and the results are consistent with previous work.[26]

Figure 1(c) presents the angular dependent LMR of sample #1 measured at $T = 2$ K by rotating the magnetic field from out-of-plane to in-plane, where $\varphi$ is the angle between the direction of magnetic field and the c-axis, as schematic illustration in Fig. 1(b). At $\varphi = 0°$, non-saturating positive LMR with quadratic *B*-field dependence is observed, as shown in the inset of Fig. 1(c), which can be attributed to the conventional compensation mechanism of the carrier density (see the Hall measurements in Fig. 3). On increasing the angle $\varphi$, the positive LMR is suppressed, and negative LMR begins to emerge at low fields with $\varphi = 32°$. For $\varphi = 90°$, i.e., the applied magnetic field is parallel to the electric current, a clearly negative LMR is appeared with the magnetic field up to 10 T, which is contrast to the previous studies,[24,26] where the negative LMR can only be observed at low fields ($B < 6$ T) at $T = 2$ K. The different transport behaviors between the ZrTe$_2$ nanoplates and bulk crystals may due to the enhancement of disorder or defect with decreasing the sample thickness[27] and the easy oxidation of sample surface of semimetallic transition metal dichalcogenides even just exposing to air a few minutes,[28] which can cause spatial fluctuations of carriers. As the temperature increases, the observed negative LMR with $B \parallel I$ decreases gradually and finally vanished above 15 K, as shown in Fig. 1(d). We also noted that the magnitude of negative LMR of the PLD grown ZrTe$_2$ thin films increases with the temperature below 10 K and then decrease with further warming,[24] this may due to the high carrier mobility in these thin films.

To get further insight into the transport properties of ZrTe$_2$ nanoplates, we carried out more measurements on samples #2 by rotating the magnetic field in the sample plane with angle $\theta$ to the current direction, as shown in Fig. 1(b). Intriguingly, at $\theta = 90°$ (i.e., the in-pane magnetic field is perpendicular to the electric current), the observed LMR exhibits a clearly negative behavior at low fields ($B < 4.5$ T) with $T = 2$ K, as shown in Fig. 2(a). When the field is further increased, the LMR turns to positive and can be well fitted by $a * B + b * B^2$, as shown in the inset of Fig. 2(b). With the titled angle $\theta$ decreases, the negative LMR shifts to high fields and can also exist up to 10 T for $\theta = 0°$, as shown in Fig. 2(a). In addition, the negative LMR for both $\theta = 90°$ and $0°$ is suppressed by increasing temperature, as shown in Figs. 2(c) and 2(d), respectively. Meanwhile, the negative LMR can only survives up to 6 K with $\theta = 90°$, where the resistivity upturn occurred (Fig. 1(a)). In contrast, when the magnetic field is applied in parallel to the electric current ($\theta = 0°$), the negative LMR disappears above 15 K, as depicted in Fig. 2(d), which is consistent with that observed in sample #1 (Fig. 1(d)).

The negative LMR was usually observed in ferromagnets due to the suppression of magnetic scatterings,[29] but for our ZrTe$_2$ nanoplates are non-magnetic. It has been predicted that ZrTe$_2$ is a topological insulator,[20-22] in which Berry curvature or axial anomaly can lead to temperature-insensitive negative LMR.[30,31] However, as shown in Figs. 1(d) and 2(d), the negative LMR observed here disappears at 20 K. Recent investigations on ZrTe$_2$ single crystals have been suggested that the isotropic negative LMR may come from the Kondo effect due to local spin of Zr atoms.[26] As shown in Fig. 2(b), after subtracting the quadratic background for $\theta = 90°$, the obtained LMR at 90° can be overlapped with the observed LMR at $\theta = 0°$ at low fields, indicating the existence of Kondo effect. Nevertheless, with the increase of magnetic field, the Kondo effect should be suppressed and finally annihilated.[26,32] Therefore, the Kondo effect could not completely explain the observed negative LMR up to 10 T for $B \parallel I$, as shown in Figs. 1(d) and 2(d). Recently, the ARPES studies have demonstrated that ZrTe$_2$ is a topological semimetal.[23,25] For topological semimetals, the chiral anomaly can lead to a negative LMR under parallel electric field and magnetic field,[33-35] which had been widely studied in various topological materials.[13-15,36-39] However, the carrier density of our ZrTe$_2$ nanoplates is about $10^{20}\ cm^{-3}$ at $T = 2$ K (Fig. 3(b)), which implies that the Fermi level is far away from the Dirac cones. Thus, the chiral anomaly hardly plays a key pole in the observed negative LMR here.

To get the underlying physics of the negative LMR in ZrTe$_2$ nanoplates, we performed the Hall measurements at different temperatures. As shown in Fig. 3(a), a nonlinear field dependence of Hall resistivity $\rho_{xy}$ is emerged below 80 K and becomes more evident as the temperature decreased. To clearly illustrate the nonlinearity of the obtained Hall resistivity at low temperature, we presented a comparison between the $\rho_{xy}$ curve and the straight solid line at 2K, as shown in the inset of Fig. 3(a). It can clearly see that the Hall resistivity deviates the linear magnetic field dependence with $B > 5T$. This nonlinear behavior signals ZrTe$_2$ is a multiband system. Herein, we use a semiclassical two-band model to fit the $\rho_{xy}$ with the following equation:[40,41]

$$\rho_{xy} = \frac{1}{e} \frac{(n_h\mu_h^2 - n_e\mu_e^2)B + (n_h - n_e)^2 \mu_h^2 \mu_e^2 B^3}{(n_h\mu_h + n_e\mu_e)^2 + (n_h - n_e)^2 \mu_h^2 \mu_e^2 B^2}, \tag{1}$$

where $e$ is the elementary electric charge, $n_h$ ($n_e$) and $\mu_h$ ($\mu_e$) are the hole (electron) carrier density and mobility, respectively. As indicated by the red dashed lines in Fig. 3(a), the Hall resistivity $\rho_{xy}$ can be well fitted by using the two-band model below 80 K. The extracted carrier densities and mobilities at different temperatures are presented in Fig. 3(b). As we can see, the carrier densities and mobilities for the holes and electrons display the same temperature-dependent tendencies, and have the same order of magnitude. The carrier density and mobility at $T = 2$ K are about $10^{20}$ cm$^{-3}$ and 300 cm$^2$Vs$^{-1}$, respectively. For a system with almost compensated electrons and holes densities, an extremely large LMR should be expected at low temperatures under the magnetic field perpendicular the sample plane.[35,42-44] However, the LMR in our ZrTe$_2$ nanoplates is about 12% at 2 K (inset of Fig. 1(c)), which could ascribe to the relatively low mobilities, as shown in Fig. 3(b).

More importantly, we find that the carrier mobility is almost unchanged below 15 K and then decreases with increasing temperature, as shown in Fig. 3(b). As mentioned above, the observed negative LMR under the parallel electromagnetic fields vanishes when the temperature is above 15 K, as shown in Figs. 1(d) and 2(d). This correlation between carrier mobility and negative LMR suggests the existence of conductivity fluctuations in the ZrTe$_2$ nanoplates, which can distort the current paths due to spatial fluctuations of the carrier mobility,[45] and leads to a negative LMR.[46,47] Since the single crystals we used for exfoliation is Zr rich,[26] such a Te deficiency or excess Zr can account for the high carrier density and the low carrier mobility in the ZrTe$_2$ nanoplates. Around the additional Zr, the scattering processes of the charge carriers are strongly affected,[45] giving rise to nonhomogeneous mobility distribution. Meanwhile, the additional Zr would be expected to primarily impact the conduction band rather than the valence band, leading to electron doping in the nanoplates, which is consistent with the measured Hall resistivity (Fig. 3(a)). It worth pointing out that, a linear LMR was observed with the in-plane magnetic field perpendicular to the electric current at high fields, as shown in Fig. 2(a), which has also been observed in ZrTe$_2$ thin films and attributed to the carrier fluctuations.[24] Thus, all the experimental evidences, lead us to believe the observed negative LMR at high fields in ZrTe$_2$ nanoplates comes from the conductivity fluctuations.

Let us turn to study the anisotropic transport properties and the related planar Hall effect (PHE) with the magnetic field consecutively rotating in the sample plane. Quantitatively, the anisotropic LMR $\rho_{xx}$ and plane Hall resistivity $\rho_{xy}^{\text{PHE}}$ can be described by the following equations:[48,49]

$$\rho_{xx} = \rho_\perp - \Delta\rho \cos^2 \theta, \tag{2}$$

$$\rho_{xy}^{PHE} = -\Delta\rho \sin\theta \cos\theta, \tag{3}$$

where $\Delta\rho = \rho_\perp - \rho_\parallel$ is the resistivity anisotropy with $\rho_\perp$ and $\rho_\parallel$ being the longitudinal resistivities for the current perpendicular to and parallel to the magnetic field, respectively. After considering the misalignment effects,[39] the obtained intrinsic results are present in Fig. 4. Here, we select two different temperatures, 2 K and 30 K, to discuss the PHE and anisotropic LMR. As clearly shown in Figs. 4(a)-4(c), both $\rho_{xy}^{PHE}$ and $\rho_{xx}$ hold a period of π, and the corresponded curves show the $\sin\theta \cos\theta$ and $\cos^2\theta$ dependence, as indicated by the fitting curves in Fig. 4(c) and the inset of Fig. 4(e), respectively. Meanwhile, we find that the magnetic field dependent PHE amplitude for both 2 K and 30 K can be well captured by a power law $B^\alpha$ with α ~ 1.85, as indicated by the red solid fitting curves in Fig. 4(e). As mentioned above, the Kondo effect holds isotropic transport properties. Therefore, the measured PHE and anisotropic LMR cannot originate from the Kondo effect. In addition, the chiral anomaly has been ruled out here, although this effect in topological semimetals can contribute to PHE.[50,51]

We also performed the temperature dependence of PHE measurements up to 250 K at $B = 12$ T. Figure 4(d) shows the extracted PHE amplitude as a function of temperature. It is clear that the PHE amplitude exhibits a non-monotonic behavior with the increase of temperature. Moreover, two extremum points are appeared, at $T = 15$ K and 60 K, above which the negative LMR (Figs. 1(d) and 2(d)) and nonlinear Hall (Fig. 3(a)) disappear, respectively. This consistency indicates the conductivity fluctuations also playing a crucial role in the PHE and anisotropic LMR.

To further investigate the effect of conductivity fluctuations, we present the parametric plots of $\rho_{xy}^{\text{PHE}}$ vs in-plane $\rho_{xx}$ with $\theta$ as a parameter under various magnetic fields at $T = 2$ K and $30$ K in Fig. 5. In strongly contrast to the concentric orbits pattern due to the chiral anomaly or the a "shock-wave" pattern contributed by the anisotropic orbital magnetoresistance,[52,53] the parametric plots in ZrTe$_2$ nanoplates exhibit ellipse-shaped pattern with the centers of orbits shifted to the right direction as increasing the magnetic fields. This special pattern may provide a strong evidence for the conductivity fluctuations.

In summary, we have investigated the negative LMR and PHE at various temperatures and magnetic fields in the ZrTe$_2$ nanoplates. Besides the Kondo effect, the correlation between the temperature dependence of negative LMR and the carrier mobility suggests the crucial role of conductivity fluctuations in the magnetotransport properties of ZrTe$_2$ nanoplates. The observation of ellipse-shaped pattern with shifted orbital center of the parametric plot provides further transport evidence for the conductivity fluctuations.

**Acknowledgements**

This research was supported by the Natural Science Foundation of China (Grants No. U19A2093, No.11774353, No.U1732274, No. U2032163, No. U2032214), the Collaborative Innovation Program of Hefei Science Center of CAS (Grant No. 2019HSC-CIP007). This research was also supported by the High Magnetic Field

Laboratory of Anhui Province. Y.W. acknowledges support from the China Postdoctoral Science Foundation Grant No. 2020M682055, the CASHIPS Director's Fund Grant No. YZJJ2021QN30.

## AUTHOR DECLARATIONS

**Conflict of Interest**

The authors have no conflicts of interests.

## DATA AVAILABILITY

The data that support the findings of this study are available from the corresponding author upon reasonable request.

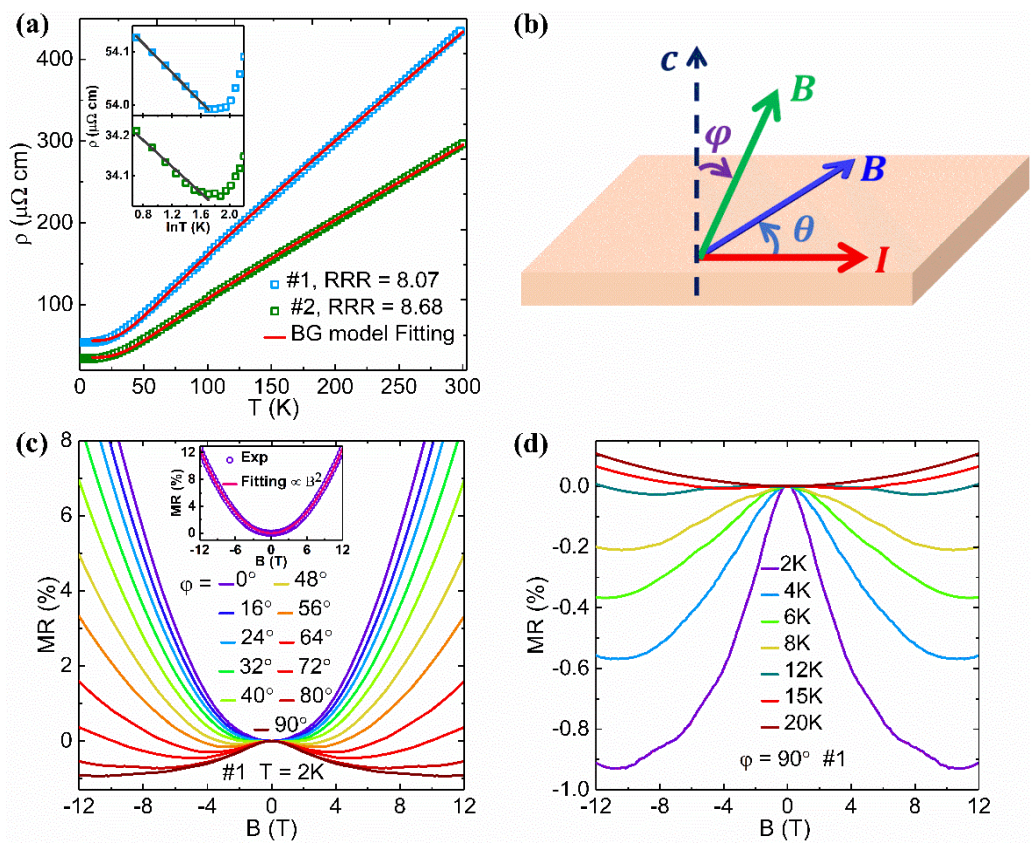

**FIG. 1.** Magnetotransport properties of ZrTe$_2$ nanoplates. (a) Temperature dependence of resistivity for samples #1 and #2. The grey curves are BG modle fitting to the risitivity between 6 and 300 K. Inset: the plot of resistivity as a logarithmic scale. (b) Schematic diagram of the experimental configuration. (c) Magnetic field denpendent LMR at different angles for #1. With increasing the titled angle, negative LMR begains to emerge and extends to high fields. Inset: The LMR at $\varphi = 0°$ can be well fitted by $B^2$. (d) The evalution of negative LMR with temperature under $\varphi = 90°$. Above 15 K, the negative LMR is disappeared.

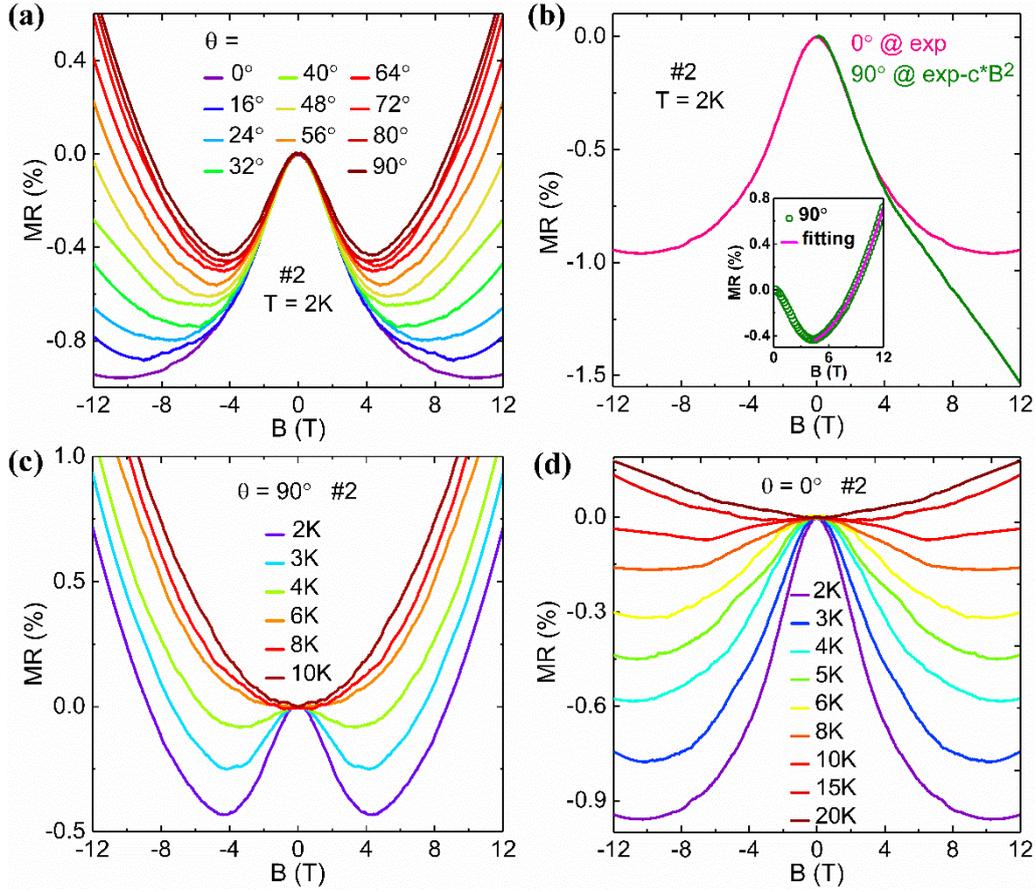

**FIG. 2.** Magnetotransport properties of sample #2 under in-plane magnetic field. (a) The magnetic dependent LMR under various angles at $T = 2$ K. The negative LMR is observed at the whole angle range. (b) The low fields LMR after subtracting the quadratic background at $\theta = 90°$ is overlapped with that measured at $\theta = 0°$. Inset: For $B > 4.5$ T, the observed LMR can be fittted by $a*B + b*B^2$ with a and b the fitting parameters. (c, d) The evalution of LMR with temperature for $\theta = 90°$ and $0°$, respectively. The negative LMR is disappeared above 6 K for $\theta = 90°$ (c), however, it is disappeared above 15 K for $\theta = 0°$ (d).

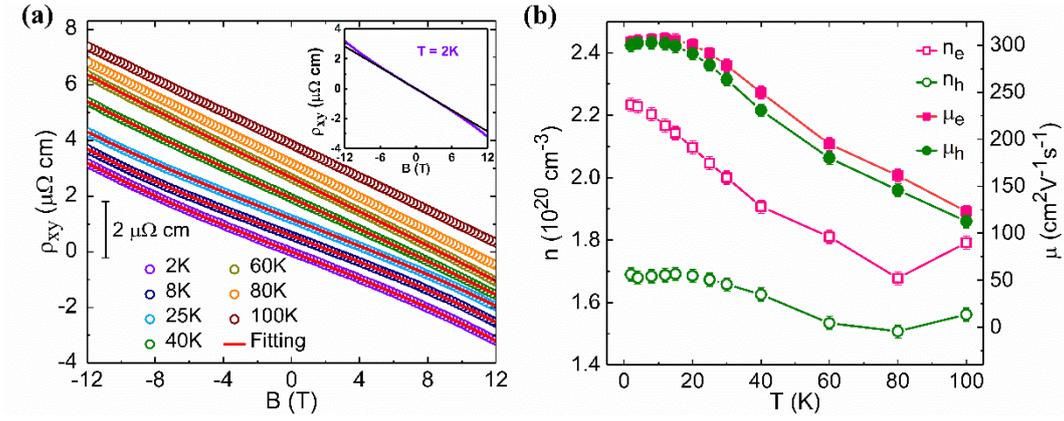

**FIG. 3.** Hall resistivity measurements on sample #1. (a) Hall resistivity as a function of magnetic field at selected temperature. For $T < 80$ K, the Hall resistivity can be well fitted with the two-band model. The data have been offset for clarity. Inset: the black solid line to illustrate the nonlinearity of Hall resistivity. (b) The carrier density and mobility at different temperatures. The fitting parameters have been constrained by the condition $\rho(B=0)^{-1} = e(n_e\mu_e + n_h\mu_h)$ with $\rho(B=0)$ the resistivity at zero magnetic field.

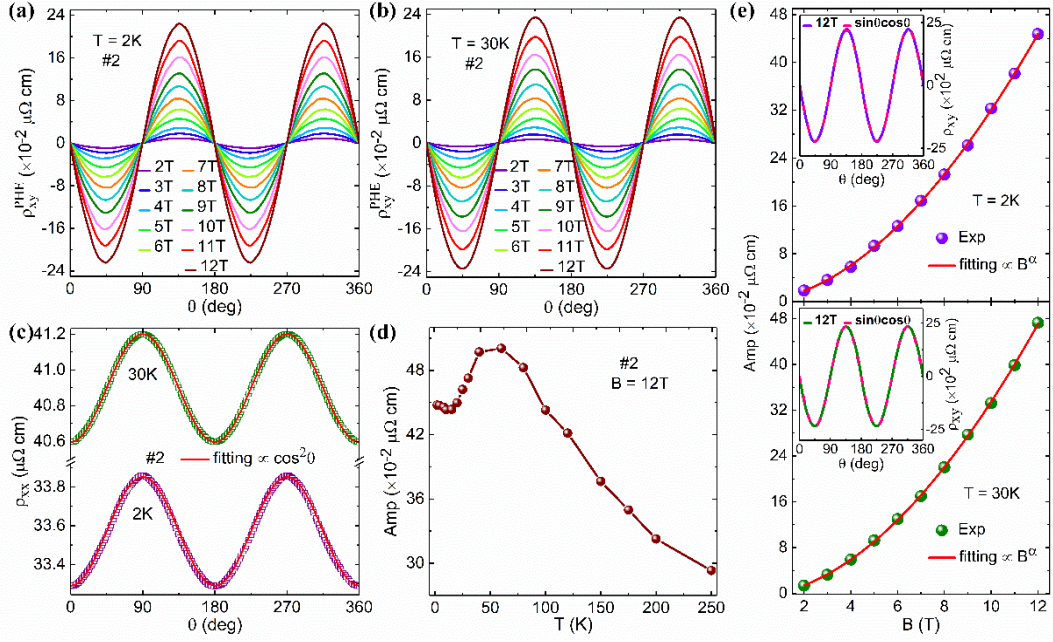

**FIG. 4.** The PHE and related anisotropic LMR measured in sample #2. (a, b) Angle dependence of the PHE at 2 K and 30 K under different magnetic fields. (c) In-plane LMR can be well fitted by Eq. (2) as indicated by the solid red curves with $B = 12$ T. (d) The amplitude of PHE as a function of temperature. With increasing temperature, two kinks (at 15 and 60 K) are observerd. (e) The magnetic field dependent amplitude of PHE at 2 K and 30 K. The solid red lines are the power law fitting curves of the amplitude of PHE, showing the field dependence with an exponent of 1.85. Inset: The fitting curves of the PHE based on Eq. (3) at 12 T.

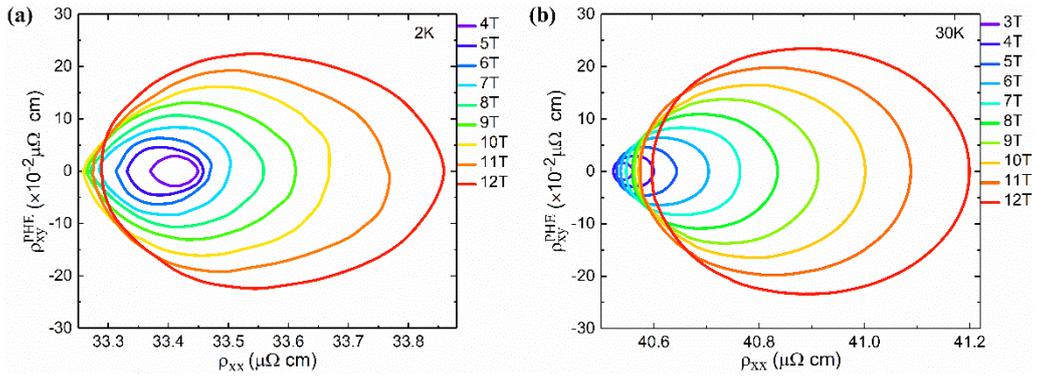

**FIG. 5.** The orbits of parametric plots at 2 K (a) and 30 K (b) at different magnetic fields. The orbits evolve to form a ellipse-shaped pattern, indicating conductivity fluctuations play an important role in the PHE measuremets.

Table 1: Competition of the atomic ratio in ZrTe$_2$ between the nanoplates and bulk crystals measured by energy-dispersive X-ray spectrometry.

| Sample | Zr Atomic present (%) | Te Atomic present (%) | Zr:Te |
|---|---|---|---|
| Nanoplate #S1 | 35.86 | 64.14 | 1.11818:2 |
| Nanoplate #S2 | 35.49 | 64.51 | 1.10029:2 |
| Nanoplate #S3 | 36.36 | 63.64 | 1.14268:2 |
| Nanoplate #S4 | 34.65 | 65.35 | 1.06044:2 |
| Bulk #S5 | 36.22 | 63.78 | 1.13578:2 |
| Buk #S6 | 35.41 | 64.59 | 1.09645:2 |
| Bulk #S7 | 35.55 | 64.45 | 1.10318:2 |